\documentclass{article}

\usepackage[table,xcdraw]{xcolor}
\usepackage{arxiv}

\usepackage[utf8]{inputenc} 
\usepackage[T1]{fontenc}    
\usepackage{hyperref}       
\usepackage{url}            
\usepackage{booktabs}       
\usepackage{amsfonts}       
\usepackage{nicefrac}       
\usepackage{microtype}      
\usepackage{lipsum}
\usepackage{graphicx}
\graphicspath{ {./images/} }
\usepackage{multirow}
\usepackage{amsthm}          
\usepackage[tbtags]{amsmath}
\usepackage{amssymb}
\usepackage{booktabs}
\usepackage{makecell}
\usepackage{algorithm}
\usepackage{algpseudocode}

\title{Amortized Bayesian Inference for Spatio-Temporal Extremes: A Copula Factor Model with Autoregression}

\author{
 Carlos A. Pasquier \\
  Programa de Posgrado en Matemática\\
  Universidad de Costa Rica\\
  San José, Costa Rica \\
  \texttt{carlos.pasquier@ucr.ac.cr} \\
   \And
 Luis A. Barboza \\
  Centro de Investigación en Matemática Pura y Aplicada\\
  Escuela de Matemática \\
  Universidad de Costa Rica\\
  San José, Costa Rica \\
  \texttt{luisalberto.barboza@ucr.ac.cr} \\
}

\begin{document}
\maketitle
\begin{abstract}
We develop a Bayesian spatio-temporal framework for extreme-value analysis that augments a hierarchical copula model with an autoregressive factor to capture residual temporal dependence in threshold exceedances. The factor can be specified as spatially varying or spatially constant, and the scale parameter incorporates scientifically relevant covariates (e.g., longitude, latitude, altitude), enabling flexible representation of geographic heterogeneity. To avoid the computational burden of the full censored likelihood, we design a Gibbs sampler that embeds amortized neural posterior estimation within each parameter block, yielding scalable inference with full posterior uncertainty for parameters, predictive quantiles, and return levels. Simulation studies demonstrate that the approach improves MCMC mixing and estimation accuracy relative to baseline specifications, particularly when using moderately more complex network architectures, while preserving heavy-tail behavior. We illustrate the methodology with daily precipitation in Guanacaste, Costa Rica, evaluating a suite of nested models and selecting the best-performing factor combination via out-of-sample diagnostics. The chosen specification reveals coherent spatial patterns in multi-year return periods and provides actionable information for infrastructure planning and climate-risk management in a tropical dry region strongly influenced by climatic factors. The proposed Gibbs scheme generalizes to other settings where parameters can be partitioned into inferentially homogeneous blocks and conditionals learned via amortized, likelihood-free methods.
\end{abstract}


\section{Introduction}

Research on natural hazards—heat waves, heavy rainfall, and windstorms—has become critical in a warming world. Evidence indicates a marked rise in the frequency of extreme events over the past five decades, underscoring the need to understand and manage these phenomena effectively \cite{pachauri2014climate}. In particular, heavy-precipitation extremes are increasing across many land regions and are projected to become more frequent and intense with additional warming; at ~4 °C of global warming, the frequency of 10-year and 50-year events is likely to double and triple, respectively \cite{IPCC_AR6_WGI_Chapter11_2021} . Recent global assessments further document unprecedented hydrological stress—record ocean heat content and sea-level rise, widespread water-related extremes, and severe regional drought–flood swings—highlighting escalating risks to people and infrastructure \cite{Fowler2021Anthropogenic}.

The increasing frequency and intensity of extremes—particularly floods and droughts—demand accurate methods for analysis and prediction, especially in vulnerable tropical regions such as Costa Rica. Observational studies across the tropics report significant shifts in precipitation extremes (e.g., higher wet‐day intensity and contributions from very wet days), with Central America and northern South America showing notable changes \cite{Aguilar2005JGR}. Global land analyses likewise indicate intensification across many tropical areas and rising annual maximum daily precipitation, consistent with broader assessments of extremes \cite{Contractor2021JClimate,IPCC_AR6_WGI_Chapter11_2021}. 

Even with abundant observational data, the core statistical difficulty remains: extreme events are rare, so parameter estimation and uncertainty quantification rely on small effective samples and heavy-tailed behavior \cite{Coles2001,DavisonPadoanRibatet2012}. This challenge is compounded by spatial complexity—geographic heterogeneity and temporal nonstationarity can distort pooling and bias inference—motivating the use of space–time extreme-value models and hierarchical frameworks that borrow strength while preserving tail-dependence structures \cite{HuserDavison2014,DavisonPadoanRibatet2012}. Within this methodological landscape, extreme-value theory offers two principal approaches: block-maxima (BM) and peaks-over-threshold (POT), with POT often preferred because it models exceedances directly rather than only maxima \cite{bucher2021horse}. Nevertheless, classical max-stable processes—the cornerstone for spatial extremes—impose a rigid dependence structure (invariant under the max operator across aggregation levels) that can contradict empirical evidence of weakening spatial dependence at higher severities; popular specifications such as the Schlather and extremal-$t$ models are also non-ergodic, and full likelihoods are tractable only in very low dimensions, making exact inference impractical in many applications \cite{HuserOpitzWadsworth2025EDS,HuserWadsworth2019JASA,GentonMaSang2011Biometrika,CastruccioHuserGenton2016}. To address these limitations, composite likelihoods assemble low-dimensional contributions into a principled surrogate for the intractable joint likelihood \cite{PadoanRibatetSisson2010JASA}, while recent likelihood-free and neural estimators aim to scale inference to higher dimensions without sacrificing fidelity in tail behavior \cite{SainsburyDaleZammitMangionHuser2024}.

Bayesian hierarchical models have proven effective in flexibly accommodating space–time structure, covariates, and latent processes under both BM and POT settings. Applications include Gaussian-process hierarchies for precipitation, spatio-temporal fire extremes, and INLA-based threshold exceedance models; recent work proposes scalable Bayesian algorithms for latent Gaussian extremes at continental scales \cite{cooley2007bayesian,TurkmanAmaralTurkmanPereira2010,OpitzHuserBakkaRue2018,CastroCamilo_Huser_Rue_2019,HazraHuserJohannesson2023}. Flexible copula constructions—particularly Gaussian and extreme-value copulas—decouple marginal tails from dependence and thus provide practical routes for modeling spatial extremal dependence beyond max-stable rigidity \cite{DavisonPadoanRibatet2012,GudendorfSegers2010}. For example, \cite{SangGelfand2009} used a Gaussian copula with GEV margins and spatial random effects, while \cite{yadav2022flexible} proposed a hierarchical copula model with relevant covariates; however, the latter assumes temporal independence and reports MCMC mixing challenges. Complementary likelihood-free studies have explored Bayesian neural estimators for spatial extremes (e.g., $r$-Pareto, inverted max-stable, random scale mixtures, conditional extremes), many do not explicitly account for temporal dependence and rich covariate structures \cite{FerreiraDeHaan2014Bernoulli,deFondeville2018,HuserWadsworth2019JASA,WadsworthTawn2022}.

In this article, we aim to develop, calibrate, and validate a Bayesian spatio–temporal extreme-value framework that (i) extends the hierarchical copula model of \cite{yadav2022flexible} by embedding an autoregressive component to capture residual temporal dependence in threshold exceedances; (ii) integrates scientifically relevant covariates (e.g., longitude, latitude, altitude, and problem-specific predictors) via a spatially varying scale; and (iii) enables scalable, likelihood-free inference through amortized Bayesian neural estimators within a Gibbs scheme, thereby providing full posterior uncertainty for parameters, predictive quantiles, and return levels.

The remainder of the article is structured as follows. Section~\ref{sec::methods} presents the proposed model and the amortized-inference estimation strategy. Section~\ref{sec::simulacion} reports a simulation study that examines the strengths and limitations of the complete estimation process. Section~\ref{sec::aplicacion} applies the framework to daily precipitation in Guanacaste, Costa Rica. Section~\ref{sec::conclusiones} concludes with the main findings and directions for future work.

\section{Statistical Methods}\label{sec::methods}

This section outlines the statistical framework and estimation strategy. We extend a spatio-temporal extreme-value model under a POT scheme, incorporate covariate-dependent scaling and an autoregressive factor for temporal dependence, and cast the model in a Bayesian hierarchical form. To avoid high-dimensional latent integration, we employ amortized inference with invertible neural networks, detail the training pipeline and a Gibbs scheme to approximate the posterior, and define validation metrics for parameter recovery and the prediction of extreme quantiles.

\subsection{Spatio-Temporal Extreme Model}
In this section, we extend the flexible factor model of \cite{yadav2022flexible} to characterize spatio‐temporal extremes under a POT (peaks-over-threshold) approach. Let  
\[
Y_t(\mathbf{s}),\quad \mathbf{s}\in S\subset\mathbb{R}^2,\quad t=1,\dots,n,
\]
denote the process observed at the finite set of locations \(\{\mathbf{s}_1,\dots,\mathbf{s}_d\}\), and collect these into  
\(\mathbf{Y}_t=(Y_t(\mathbf{s}_1),\dots,Y_t(\mathbf{s}_d))^\top\). We assume \(\{\mathbf{Y}_t\}\) are i.i.d.\ replicates of a base process \(Y(\mathbf{s})\) and factorize  
\[
Y_t(\mathbf{s}) = \alpha(\mathbf{s})\,X_{1t}(\mathbf{s})\,X_{2t}(\mathbf{s})\,X_{3t}(\mathbf{s}),
\]
where:
\begin{itemize}
	\item \(X_{1t}(\mathbf{s})\) is spatial white noise (i.i.d.\ across sites, unit mean, Weibull‐tailed distribution \(F_1\)):
	\[X_{1tj} = \frac{E_{1tj}^{\beta_1}}{\Gamma(1+\beta_1)}, \;E_{1tj} \overset{\text{iid}}{\sim}   \text{Exp}(1); \quad j = 1, \dots, d; \quad t = 1, \dots, n; \quad \beta_1 > 0,  \]
	\item \(X_{2t}(\mathbf{s})\equiv X_{2t}\) is a spatially constant factor (i.i.d.\ in time, unit mean, Weibull‐tailed \(F_2\));
	\[X_{2t} = \frac{E_{2t}^{\beta_2}}{\Gamma(1+\beta_2)}, \;E_{2t} \overset{\text{i.i.d}}{\sim}   \text{Exp}(1); \quad t = 1, \dots, n; \quad \beta_2 > 0,  \]
	\item \(X_{3t}(\mathbf{s})\) is a nontrivial spatial process with copula \(C_{X_3}\) and regularly varying margins \(F_3\). Specifically, we define the marginal distribution $F_3$ as an inverse-gamma (IG) with shape $\beta_3>1$ and scale $\beta_3-1$, which ensures $\mathbb{E}(X_{3tj})=1$. We model the copula $C_{\mathbf{X}_3}$ as Gaussian with an exponential correlation function $\rho(h)=\exp(-h/\rho)$ for $h\ge 0$ and range parameter $\rho>0$. This specification yields unit-mean heavy-tailed margins together with an exponentially decaying spatial dependence structure. 
	\item \(\alpha(\mathbf{s})=\exp\bigl(\gamma_0\mathbf{1}_d+\sum_{k=1}^p\gamma_k\mathbf{Z}_k\bigr)\) links spatial covariates \(\mathbf{Z}_k\) (e.g.\ latitude, longitude, altitude) via a log‐linear regression.
\end{itemize}
Applying Breiman’s lemma \cite{Breiman1965ArcSin} to this multiplicative construction shows that the product process is regularly varying whenever at least one factor is regularly varying and the remaining factors are light-tailed. Consequently, the composite process inherits heavy-tailed behavior even if some marginal components are not themselves heavy-tailed.

To capture residual temporal dependence in threshold exceedances, we replace \(X_{2t}\) by an AR(1) factor \(X_{2t}^{\mathrm{AR}}\):
\[
\log X_{2t}^{\mathrm{AR}}
=(1-\phi)\tau + \phi\,\log X_{2,(t-1)}^{\mathrm{AR}} + \varepsilon_t,\quad 
\varepsilon_t\sim N(0,\sigma^2),
\]
with \(\tau\) chosen so that \(\mathbb{E}[X_{2t}^{\mathrm{AR}}]=1\).  
To probe the role of spatial dependence, we consider variants that control whether the autoregressive factor $\mathbf{X}_{2t}^{\mathrm{AR}}$ and the idiosyncratic factor $\mathbf{X}_{1t}$ are spatially constant. A factor is spatially constant when all site-specific components coincide, i.e., $X^{\mathrm{AR}}_{2t1}=X^{\mathrm{AR}}_{2t2}=\cdots=X^{\mathrm{AR}}_{2td}$. We denote such cases with the superscript $c$, writing $\mathbf{X}_{2t}^{\mathrm{AR}\text{-}c}$ and $\mathbf{X}_{1t}^{c}$.

Our proposed model is
\begin{align}
	\mathbf{Y}_t \;=\; \boldsymbol{\alpha}\;\mathbf{X}_{1t}\;\mathbf{X}_{2t}^{\mathrm{AR}}\;\mathbf{X}_{3t}, 
	\qquad t=1,\dots,n,
	\label{eq:modelo_propuesto}
\end{align}
which jointly captures flexible marginal tails, spatial dependence, covariate effects, and temporal correlation in threshold exceedances. In our comparisons, we specify $\mathbf{X}_{2t}^{\mathrm{AR}}$ and $\mathbf{X}_{1t}$ either as spatially constant (superscript $c$) or spatially varying across sites, yielding a set of nested scenarios for assessing sensitivity to spatial structure.  For technical details on the general construction of multiplicative extreme‐value models and the analysis of their tail behavior, we direct readers to \cite{yadav2022flexible}, \cite{yadav2021spatial}, and \cite{breiman1965limit}.

\subsection{Estimation}
To enable estimation of model~\eqref{eq:modelo_propuesto}, we reformulate its joint distribution in hierarchical form, which facilitates Bayesian inference and modular specification of conditional components:
\begin{align}
	Y_{tj}\mid X_{2t}^{\mathrm{AR}},X_{3t},\Theta_{X_1},\Theta_{\alpha}
	&\;\overset{\mathrm{ind}}{\sim}\;
	F_1\!\bigl(\cdot \mid \alpha_j\,X_{2t,j}^{\mathrm{AR}}\,X_{3t,j};\,\Theta_{X_1}\bigr),\notag \\
	X_{2t}^{\mathrm{AR}}\mid \Theta_{X_2}
	&\;\sim\;F_2\!\bigl(\cdot;\,\Theta_{X_2}\bigr),\notag \\
	X_{3t}\mid \Theta_{X_3}^{\mathrm{mar}},\Theta_{X_3}^{\mathrm{dep}}
	&\;\overset{\mathrm{ind}}{\sim}\;
	C_{X_3}\Bigl(F_3(\cdot;\Theta_{X_3}^{\mathrm{mar}}),\ldots; \,\Theta_{X_3}^{\mathrm{dep}}\Bigr),\notag \\
	\Theta&\;\sim\;\pi(\Theta),
	\label{eq:modelo_sim}
\end{align}
where 
\(\Theta=(\Theta_{\alpha}^T,\Theta_{X_1}^T,\Theta_{X_2}^T,\Theta_{X_3}^{\mathrm{mar}\,T},\Theta_{X_3}^{\mathrm{dep}\,T})^T\), $\Theta_{\alpha}=(\gamma_k)_{k=1}^p$, $\Theta_{X_1}=\beta_1$, $\Theta_{X_2}=(\phi,\sigma)$, $\Theta^{\text{mar}}_{X_3} = \beta_3$ and $\Theta^{\text{dep}}_{X_3} = \rho$.
We treat \(\{X_{2t}^{\mathrm{AR}}\}_{t=1}^n\) (dimension \(n\)) and \(\{X_{3t}\}_{t=1}^n\) (dimension \(nd\)) as latent variables.  The joint posterior factorizes as
\[
\pi(\Theta,X_{2}^{\mathrm{AR}},X_{3}\mid Y)
\;\propto\;
\pi(Y\mid X_{2}^{\mathrm{AR}},X_{3},\Theta_{X_1})\,
\pi(X_{2}^{\mathrm{AR}}\mid\Theta_{X_2})\,
\pi(X_{3}\mid\Theta_{X_3})\,
\pi(\Theta),
\]
and we recover the posterior distribution as:
\[
\pi(\Theta\mid Y)
=\int\!\!\int \pi(\Theta,X_{2}^{\mathrm{AR}},X_{3}\mid Y)\,dX_{2}^{\mathrm{AR}}\,dX_{3}.
\]
Because this integral is high‐dimensional, we implement amortized inference: we train a first set of networks
\(\,R_\alpha\) to approximate the posterior of \(\Theta_{\alpha}\) and a second set of networks
\(\,R_X\) to approximate the joint posterior of \(\Theta_{X_2}\) and \(\Theta_{X_3}\), 
thereby avoiding costly MCMC over all latent variables simultaneously. In the following sections, we provide a concise overview of the estimation methodology.
 
\subsubsection{Amortized Inference}
\label{sec:amortinf}

We address the limitations of MCMC—particularly its slow convergence in high‐dimensional parameter spaces—by adopting a Bayesian neural‐network framework that explicitly models parameter uncertainty and improves generalization \cite{radev2020bayesflow}. In particular, BayesFlow \cite{radev2023jana} implements globally amortized inference via a conditional invertible neural network (cINN) \(f_{\boldsymbol{\phi}}\), which learns a bijective mapping between latent Gaussian variables \(\boldsymbol{z}\sim\mathcal{N}(0,I)\) and model parameters \(\boldsymbol{\Theta}\) conditioned on observations \(\boldsymbol{y}\). We train \(f_{\boldsymbol{\phi}}\) by minimizing the expected Kullback–Leibler divergence  
\[
\mathbb{E}_{p(\boldsymbol{y},\boldsymbol{\Theta})}\bigl[\tfrac12\|f_{\boldsymbol{\phi}}(\boldsymbol{\Theta};\boldsymbol{y})\|^2 - \log\bigl|\det J_{f_{\boldsymbol{\phi}}}\bigr|\bigr],
\]  
using Monte Carlo samples \(\{(\boldsymbol{y}^{(m)},\boldsymbol{\Theta}^{(m)})\}\) from model \eqref{eq:modelo_sim}. To handle variable‐size datasets, we introduce a summary network \(h_{\boldsymbol{\psi}}(\boldsymbol{y}_{1:n})\) that learns informative statistics directly from the data, replacing hand‐crafted summaries. We jointly optimize \((\boldsymbol{\phi},\boldsymbol{\psi})\) via stochastic gradient descent on the loss  
\begin{align}
\mathcal{L}(\boldsymbol{\phi},\boldsymbol{\psi})
=\frac{1}{M}\sum_{m=1}^M\Bigl[\tfrac12\|f_{\boldsymbol{\phi}}(\boldsymbol{\Theta}^{(m)};h_{\boldsymbol{\psi}}(\boldsymbol{y}_{1:n}^{(m)}))\|^2
-\log\bigl|\det J_{f_{\boldsymbol{\phi}}}\bigr|\Bigr].	
\label{eq:montecarlo_completo}
\end{align}
Under perfect convergence, the trained cINN and summary network yield exact posterior samples. To illustrate this procedure, we present Algorithm~\ref{alg:bayesflow}, which outlines the essential steps of the BayesFlow framework.

\subsubsection{Architectures}
We organize the realized values \(y_{tj}=Y_t(\mathbf{s}_j)\) of the dependent variable of interest into an \(n\times d\) data matrix 
\[
\mathbf{Y}=\bigl[y_{tj}\bigr]_{t=1,\dots,n;\,j=1,\dots,d},
\]
then apply site‐specific thresholds \(u_j\) to obtain the censored matrix \(\mathbf{Y}^\vee\) with entries: 
\begin{align}
y_{tj}^\vee=\max\{y_{tj},u_j\}.
\label{eq::truncar}	
\end{align}

For the covariate‐scale summary network \(R_{\alpha}\), we concatenate \(\mathbf{Y}^\vee\) with the hyperparameter values \(\Theta_{X_2^{\mathrm{AR}}},\Theta_{X_3}^{\mathrm{mar}},\Theta_{X_3}^{\mathrm{dep}}\) as additional columns.  We feed this \([n\times(d+4)]\) tensor into two stacked LSTM layers—first producing an \([n\times n_{\mathrm{LSTM}}]\) sequence, then reducing to a single \([n_{\mathrm{LSTM}}]\) vector—followed by two dense layers (ReLU then ELU) to yield a fixed‐length ($n_\mathrm{Dense}$) summary for inferring \(\Theta_{\alpha}\).

For the latent‐factor network \(R_{X}\), we compute:
\begin{align}
	x_{tj}^\vee=\frac{y_{tj}^\vee}{\alpha_j},
	\label{eq::truncar2}
\end{align}
then we reshape each time slice into a \(d_1\times d_2\) grid such that $d_1\cdot d_2=d$, and pack into a \([n,d_1,d_2,1]\) tensor.  We apply two TimeDistributed 2D convolutions (3×3 kernels, 32 and 64 filters), flatten the activations to \([n,d_1d_2\cdot64]\), and process them through two LSTMs and two dense layers (ReLU, ELU) to obtain summary statistics for \(\Theta_{X_2^{\mathrm{AR}}}\) and \(\Theta_{X_3}\).  

This architecture balances expressivity and efficiency for moderate grid sizes (\(d_1,d_2\approx5\)–10).  Table \ref{tab::archiqs} in the appendix contains a summary of the architecture details.

\subsubsection{Proposed Algorithms}
As described in \ref{sec:amortinf}, BayesFlow trains summary networks by drawing simulations from the prior predictive distribution to produce posterior samples of the hyperparameters. Algorithm~\ref{alg:pasquier1} details the training pipeline for the networks \(R_{\alpha}\) and \(R_{X}\). In particular, steps 8 and 20 simulate the censored observations \(y_{tj}^\vee\) and \(x_{tj}^\vee\) according to equations \eqref{eq::truncar} and \eqref{eq::truncar2} respectively. Importantly, when training \(R_{X}\), we omit sampling the covariate‐scale hyperparameters \(\Theta_{\alpha}\), since they do not affect the generation of \(x_{tj}^\vee\).

\begin{algorithm}
	\caption{Amortized Bayesian Inference via BayesFlow for Model~\eqref{eq:modelo_propuesto}}\label{alg:pasquier1}
	\begin{algorithmic}[1]
		\State $n$ is the number of process observations.
		\State $\boldsymbol{\Theta}=(\boldsymbol{\Theta}_{\alpha}^T,\;\boldsymbol{\Theta}_{X_1}^T,\;\boldsymbol{\Theta}_{X_{2t}^{\mathrm{AR}}}^T,\;\boldsymbol{\Theta}_{X_3}^{\mathrm{mar}\,T},\;\boldsymbol{\Theta}_{X_3}^{\mathrm{dep}\,T})^T$.
		\State $\boldsymbol{\Theta}_X=(\boldsymbol{\Theta}_{X_{2t}^{\mathrm{AR}}}^T,\;\boldsymbol{\Theta}_{X_3}^{\mathrm{mar}\,T},\;\boldsymbol{\Theta}_{X_3}^{\mathrm{dep}\,T})^T$.
		
		\Statex
		\State \textbf{Training Phase for $R_{\alpha}$} (online learning, batch size $M$):
		\Repeat
		\For{$m = 1,\dots,M$}
		\State Sample full parameter vector: $\boldsymbol{\Theta}^{(m)}\sim p(\boldsymbol{\Theta})$
		\State Simulate censored observations 
		\[
		\boldsymbol{y}_{1:n}^{(m)}=\bigl\{y_{tj}^\vee\bigr\}_{t=1,\dots,n}^{j=1,\dots,d}
		\]
		from the prior $\pi(\boldsymbol{\Theta})$.
		\State Compute summary: $\tilde{\boldsymbol{y}}^{(m)}=h_{\psi}(\boldsymbol{y}_{1:n}^{(m)})$
		\State Forward‐pass: $\boldsymbol{z}^{(m)}=f_{\phi}(\boldsymbol{\Theta}_{\alpha}^{(m)};\,\tilde{\boldsymbol{y}}^{(m)})$
		\EndFor
		\State Compute loss via \eqref{eq:montecarlo_completo} using batch 
		$\{(\boldsymbol{\Theta}_{\alpha}^{(m)},\tilde{\boldsymbol{y}}^{(m)},\boldsymbol{z}^{(m)})\}_{m=1}^M$
		\State Update $\phi,\psi$ by backpropagation
		\Until{convergence to $\hat\phi,\hat\psi$}
		
		\Statex
		\State \textbf{Training Phase for $R_{X}$} (online learning, batch size $M$):
		\Repeat
		\For{$m = 1,\dots,M$}
		\State Sample latent‐factor parameters: $\boldsymbol{\Theta}_X^{(m)}\sim p(\boldsymbol{\Theta}_X)$
		\State Simulate censored factors
		\[
		\boldsymbol{x}_{1:n}^{(m)}=\bigl\{x_{tj}^\vee\bigr\}_{t=1,\dots,n}^{j=1,\dots,d}
		\]
		from the prior $\pi(\boldsymbol{\Theta}_X)$.
		\State Compute summary: $\tilde{\boldsymbol{x}}^{(m)}=h_{\psi}(\boldsymbol{x}_{1:n}^{(m)})$
		\State Forward‐pass: $\boldsymbol{z}^{(m)}=f_{\phi}(\boldsymbol{\Theta}_X^{(m)};\,\tilde{\boldsymbol{x}}^{(m)})$
		\EndFor
		\State Compute loss via \eqref{eq:montecarlo_completo} using batch 
		$\{(\boldsymbol{\Theta}_X^{(m)},\tilde{\boldsymbol{x}}^{(m)},\boldsymbol{z}^{(m)})\}_{m=1}^M$
		\State Update $\phi,\psi$ by backpropagation
		\Until{convergence to $\hat\phi,\hat\psi$}
	\end{algorithmic}
\end{algorithm}

After training the summary networks, we use \(R_{\alpha}\) to draw posterior samples of the covariate‐scale parameters \(\boldsymbol{\Theta}_{\alpha}\) conditional on the latent‐factor parameters \(\boldsymbol{\Theta}_{X}\), and we use \(R_{X}\) to sample \(\boldsymbol{\Theta}_{X_2^{\mathrm{AR}}}\) and \(\boldsymbol{\Theta}_{X_3}^{\mathrm{mar}}, \boldsymbol{\Theta}_{X_3}^{\mathrm{dep}}\), conditional on the sample parameters $\boldsymbol{\Theta}_{\alpha}$. Algorithm~\ref{alg:pasquier2} then interleaves these conditional draws within a Gibbs sampler to generate joint posterior samples for the proposed model in \eqref{eq:modelo_propuesto}.  

\begin{algorithm}
	\caption{Gibbs Sampling with Amortized Inference}\label{alg:pasquier2}
	\begin{algorithmic}[1]
		\State \textbf{Input:} Summary networks $R_{\alpha}$ and $R_{X}$; number of iterations $n_{\mathrm{iter}}$; censored observations $y_{tj}^\vee$ for $t=1,\ldots,n$, $j=1,\ldots,d$.
		\State Initialize latent‐factor hyperparameters 
		\[
		\boldsymbol{\Theta}_X^{(0)}
		= \bigl(\boldsymbol{\Theta}_{X_{2t}^{\mathrm{AR}}}^{(0)},\,\boldsymbol{\Theta}_{X_3}^{\mathrm{mar}(0)},\,\boldsymbol{\Theta}_{X_3}^{\mathrm{dep}(0)}\bigr)^T.
		\]
		\For{$i = 0,1,\dots,n_{\mathrm{iter}}-1$}
		\State Use $R_{\alpha}$ with inputs $\{y_{tj}^\vee\}$ and $\boldsymbol{\Theta}_X^{(i)}$ to draw posterior samples $\boldsymbol{\Theta}_{\alpha}^{(i)}$.
		\State Compute scale factors $\alpha_j$ from $\boldsymbol{\Theta}_{\alpha}^{(i)}$ and form censored factors 
		\[
		x_{tj}^\vee = \frac{y_{tj}^\vee}{\alpha_j}.
		\]
		\State Use $R_{X}$ with inputs $\{x_{tj}^\vee\}$ to draw posterior samples $\boldsymbol{\Theta}_X^{(i+1)}$.
		\EndFor
		\State \Return Posterior samples of the full parameter vector 
		\(\boldsymbol{\Theta}=(\boldsymbol{\Theta}_{\alpha}^T,\boldsymbol{\Theta}_{X_1}^T,\boldsymbol{\Theta}_{X_{2t}^{\mathrm{AR}}}^T,\boldsymbol{\Theta}_{X_3}^{\mathrm{mar}T},\boldsymbol{\Theta}_{X_3}^{\mathrm{dep}T})^T.\)
	\end{algorithmic}
\end{algorithm}

\subsection{Model Performance Evaluation}\label{sec:metrics}

After applying the above algorithms to the specific case studies presented below, we evaluate the model’s performance using both parameter‐recovery and extreme‐value metrics:

\begin{itemize}
	\item \textbf{Absolute bias (AB):} for each true parameter \(\Theta^{\mathrm{true}}\) and posterior draws \(\{\Theta^{(i)}\}_{i=1}^{N}\), we compute
	\[
	\mathrm{AB}(\Theta)
	= \bigl|\tfrac{1}{N}\sum_{i=1}^N\bigl(\Theta^{(i)}-\Theta^{\mathrm{true}}\bigr)\bigr|.
	\]

	\item \textbf{Posterior standard error (SE):} we estimate
	\[
	\mathrm{SE}(\Theta)
	= \sqrt{\tfrac{1}{N-1}\sum_{i=1}^N\bigl(\Theta^{(i)}-\bar\Theta\bigr)^2},
	\quad \bar\Theta=\tfrac{1}{N}\sum_{i=1}^N\Theta^{(i)}.
	\]
	
	\item \textbf{Effective sample size per minute (ESS/min):} Compute the effective sample size
	
	$$
	\mathrm{ESS} \;=\; \frac{N}{\,1+2\sum_{k=1}^{K}\rho_k\,},
	$$
	
	where $N$ is the posterior sample size, $\rho_k$ is the lag-$k$ autocorrelation of the posterior draws, and $K$ is the largest lag at which $\rho_k$ is non-negligible. Scale by the runtime (minutes) to obtain
	
	$$
	\mathrm{ESS/min} \;=\; \frac{\mathrm{ESS}}{\text{time}_{\mathrm{min}}}.
	$$
	
	\item \textbf{95\% credible interval (CI):} we report the 2.5\% and 97.5\% quantiles of a given posterior sample:
	\(\bigl[\Theta_{(0.025)},\Theta_{(0.975)}\bigr]\).
	
	\item \textbf{Coefficient of determination ($R^2$).} After training the two networks in Algorithm \ref{alg:pasquier1}, we assess inferential accuracy by selecting $R$ random ground-truth parameter settings $\{\Theta_r^{\mathrm{true}}\}_{r=1}^R$. For each setting $r$, we draw a posterior sample $\{\Theta_r^{(i)}\}_{i=1}^{N}$ and compute its posterior mean $\bar{\Theta}_r = \frac{1}{N}\sum_{i=1}^{N}\Theta_r^{(i)}$. We then summarize performance with
	
	$$
	R^2 \;=\; 1 \;-\; 
	\frac{\sum_{r=1}^{R}\!\left(\bar{\Theta}_r - \Theta_r^{\mathrm{true}}\right)^2}
	{\sum_{r=1}^{R}\!\left(\Theta_r^{\mathrm{true}} - \overline{\Theta^{\mathrm{true}}}\right)^2},
	\qquad 
	\overline{\Theta^{\mathrm{true}}}=\frac{1}{R}\sum_{r=1}^{R}\Theta_r^{\mathrm{true}}.
	$$
\end{itemize}

To assess predictive accuracy for extreme quantiles while accounting for posterior uncertainty, we use:

\begin{itemize}
	\item \textbf{Mean Quantile Absolute Error (MQAE).}
	
	$$
	\mathrm{MQAE} \;=\; \frac{1}{(99-c_u)\, n\, d\, N}
	\sum_{i=1}^{N}\sum_{c=c_u}^{99}\sum_{t=1}^{n}\sum_{j=1}^{d}
	\bigl|\, q_{tj}^{(c,\mathrm{obs})} - q_{tj}^{(c,i)} \,\bigr|,
	$$
	
	where $q_{tj}^{(c,\mathrm{obs})}$ is the empirical $c\%$ quantile of the observed data at time $t$ and site $j$, and $q_{tj}^{(c,i)}$ is the corresponding $c\%$ quantile of the process simulated under posterior draw $\Theta^{(i)}$. Here $c_u\in[0,99)$ is the lower quantile index from which errors are evaluated (e.g., $c_u=75$), $n$ is the number of time points, $d$ the number of sites, and $N$ the number of posterior draws. Smaller values indicate better upper-tail quantile accuracy.

	\item \textbf{Mean Quantile Squared Error (MQSE):}
	\[
	\mathrm{MQSE}
	= \frac{1}{(99-c_u)\,n\,d\,N}
	\sum_{i=1}^N \sum_{c=c_u}^{99} \sum_{t=1}^n \sum_{j=1}^d
	\bigl(q_{tj}^{(c,\mathrm{obs})} - q_{tj}^{(c,i)}\bigr)^2.
	\]
\end{itemize}

\section{Simulation Study}\label{sec::simulacion}

In this section, we evaluate the Gibbs–sampling estimator (Algorithm~\ref{alg:pasquier2}) on synthetic data generated from model~\eqref{eq:modelo_propuesto}. To enable comparison with prior work, we follow \cite{yadav2022flexible} when specifying the spatial grid, latent factors, correlation structure, and parameter values. We simulate \(n=200\) temporal replicates on \(d=100\) locations arranged on a uniform \([0,1]^2\) grid. We draw the latent factor \(X_{3t}\) from a Gaussian copula with isotropic exponential correlation \(\rho(h)=\exp(-\|h\|/\rho)\) and range \(\rho=0.5\), and we censor each site at its empirical 75th percentile. The true parameters are \(\phi=0.7\), \(\sigma=1\), \(\beta_3=5\), and \(\rho=0.5\). We specify the spatial scale as
\[
\boldsymbol{\alpha}=\exp\bigl(\gamma_0\mathbf{1}_d + \gamma_1 Z_1 + \gamma_2 Z_2 + \gamma_3 Z_3\bigr),
\]
with \(\gamma_0=e^1\) and \(\gamma_1=\gamma_2=\gamma_3=1\); here \(Z_1, Z_2\) are Cartesian coordinates and \(Z_3\sim N(0,1)\) is an independent covariate.

Table~\ref{tab:arquitecturas_modelos} compares five summary‐network architectures by varying the number of LSTM units \(n_{\mathrm{LSTM}}\) and dense neurons \(n_{\mathrm{Dense}}\) for both \(R_X\) and \(R_{\alpha}\). This design quantifies how architectural capacity influences estimation accuracy and computational cost: increasing \(n_{\mathrm{LSTM}}\) targets temporal dependence, whereas increasing \(n_{\mathrm{Dense}}\) targets nonlinear mappings from summaries to parameters. The scenarios span simple to high‐capacity configurations, enabling a comprehensive assessment of accuracy–efficiency trade‐offs and informed model selection under resource constraints.
\begin{table}
	\centering
	\begin{tabular}{c|c|c}
		\toprule
		\textbf{Scenarios} & $n_{\mathrm{LSTM}}$ & $n_{\mathrm{Dense}}$ \\
		\midrule
		1 & 128  & 128  \\ 
		2 & 1024 & 128  \\
		3 & 128  & 1024 \\ 
		4 & 1024 & 1024 \\
		5 & 1000 & 2000 \\ 
		\bottomrule
	\end{tabular}
	\caption{Architectural configurations for each scenario: number of LSTM units and number of dense neurons.}
	\label{tab:arquitecturas_modelos}
\end{table}

We place weakly informative priors to ensure numerical stability and avoid degenerate latent processes:
\[
\phi\sim\mathrm{Uniform}(-0.85,0.85),\quad
\sigma\sim\mathrm{Uniform}(0.05,3),\quad
\beta_3\sim\mathrm{Uniform}(2,15),\quad
\rho\sim\mathrm{Uniform}(0,2\delta),\quad
\gamma_i\sim N(0,2),
\]
where \(\delta\) is the maximum intersite Euclidean distance. These ranges prevent \(X_{2}\) and \(X_{3}\) from collapsing to trivial behavior and promote stable training.

Table~\ref{tab:simulacion_p} reports performance under the metrics from Section~\ref{sec:metrics}. Scenario~2 attains the smallest absolute bias (AB) for most parameters, indicating the highest pointwise accuracy, while Scenario~3 shows larger AB for \(\gamma_0,\phi,\sigma,\rho\). Posterior standard errors (SE) are broadly comparable across scenarios; Scenarios~3 and~5 achieve the lowest SE for \(\sigma,\beta_3,\rho\). Scenarios~3 and~5 also yield the narrowest 95\% credible‐interval widths—especially for \(\gamma_0,\sigma,\beta_3\)—whereas Scenario~2 exhibits the widest interval for \(\beta_3\) (6.19). All configurations deliver high ESS/min, indicating efficient sampling with weak autocorrelation. The coefficient of determination reaches \(\approx 0.99\) for \(\gamma_0\)–\(\gamma_3\); Scenario~5 achieves the largest \(R^2\) for \(\phi\) (0.83). In contrast, all scenarios obtain lower \(R^2\) for \(\rho\), with Scenario~5 lowest (0.25), suggesting that the spatial‐range parameter remains challenging to recover accurately.

Figures~\ref{fig:traceplot_simulacion} and~\ref{fig:histograma_simulacion} (appendix) display posterior trace plots and histograms for Scenarios~1 and~5. Increasing network capacity improves mixing and posterior accuracy for \(\gamma_3,\phi,\rho,\beta_3\), with limited gains for \(\gamma_0,\gamma_1,\gamma_2\). Overall, the chains mix well, with only mild posterior bias relative to the ground truth.

\begin{table}
	\centering
	\begin{tabular}{c|c|c|c|c|c|c|c|c|c}
		\toprule
		& \textbf{Scenarios} & \textbf{$\gamma_0$} & \textbf{$\gamma_1$} & \textbf{$\gamma_2$} & \textbf{$\gamma_3$} & \textbf{$\phi$} & \textbf{$\sigma$} & \textbf{$\beta_3$} & \textbf{$\rho$} \\
		\midrule
		\multirow{5}{*}{\textbf{AB}}
		& 1 & 0.12 & 0.01 & 0.06 & 0.10 & 0.04 & 0.01 & 0.62 & 0.07 \\
		& 2 & 0.10 & 0.01 & 0.10 & 0.00 & 0.08 & 0.04 & 0.22 & 0.01 \\
		& 3 & 0.21 & 0.03 & 0.13 & 0.09 & 0.20 & 0.10 & 0.34 & 0.20 \\
		& 4 & 0.17 & 0.01 & 0.10 & 0.14 & 0.09 & 0.06 & 0.36 & 0.02 \\
		& 5 & 0.22 & 0.04 & 0.09 & 0.02 & 0.02 & 0.01 & 0.25 & 0.04 \\
		\midrule
		\multirow{5}{*}{\textbf{SE}}
		& 1 & 0.23 & 0.10 & 0.10 & 0.13 & 0.09 & 0.07 & 1.02 & 0.19 \\
		& 2 & 0.22 & 0.09 & 0.09 & 0.12 & 0.09 & 0.07 & 1.52 & 0.21 \\
		& 3 & 0.19 & 0.08 & 0.09 & 0.12 & 0.07 & 0.09 & 0.69 & 0.16 \\
		& 4 & 0.22 & 0.08 & 0.09 & 0.13 & 0.09 & 0.09 & 1.22 & 0.20 \\
		& 5 & 0.22 & 0.09 & 0.08 & 0.12 & 0.09 & 0.12 & 0.67 & 0.19 \\
		\midrule
		\multirow{5}{*}{\textbf{95\% CI Width}}
		& 1 & 0.89 & 0.39 & 0.38 & 0.52 & 0.37 & 0.28 & 4.09 & 0.72 \\
		& 2 & 0.87 & 0.36 & 0.37 & 0.45 & 0.34 & 0.31 & 6.19 & 0.77 \\
		& 3 & 0.76 & 0.32 & 0.34 & 0.45 & 0.30 & 0.38 & 2.76 & 0.59 \\
		& 4 & 0.86 & 0.33 & 0.36 & 0.49 & 0.38 & 0.40 & 4.66 & 0.72 \\
		& 5 & 0.89 & 0.35 & 0.31 & 0.48 & 0.37 & 0.47 & 2.79 & 0.68 \\
		\midrule
		\multirow{5}{*}{\textbf{ESS/min}}
		& 1 &  96 & 105 & 112 & 104 & 106 & 108 & 102 & 108 \\
		& 2 &  96 & 100 & 110 & 111 & 114 & 114 & 110 & 110 \\
		& 3 & 106 & 106 & 112 & 108 & 110 & 112 & 111 & 111 \\
		& 4 & 105 & 109 & 111 & 110 & 110 & 104 & 107 & 109 \\
		& 5 & 103 & 102 & 109 & 110 & 110 & 113 & 101 & 112 \\
		\midrule
		\multirow{5}{*}{\textbf{$R^2$}}
		& 1 & 0.98 & 0.99 & 0.99 & 0.98 & 0.75 & 0.94 & 0.56 & 0.41 \\
		& 2 & 0.97 & 0.99 & 0.99 & 0.98 & 0.74 & 0.94 & 0.59 & 0.39 \\
		& 3 & 0.97 & 0.99 & 0.99 & 0.98 & 0.80 & 0.95 & 0.54 & 0.35 \\
		& 4 & 0.96 & 0.99 & 0.99 & 0.98 & 0.81 & 0.97 & 0.61 & 0.37 \\
		& 5 & 0.97 & 0.99 & 0.99 & 0.98 & 0.83 & 0.95 & 0.53 & 0.25 \\
		\bottomrule
	\end{tabular}
	\caption{Performance metrics for each scenario: absolute bias (AB), posterior standard error (SE), 95\% credible‐interval width, effective sample size per minute (ESS/min), and coefficient of determination (\(R^2\)).}
	\label{tab:simulacion_p}
\end{table}

All training runs and simulations reported in this and the following sections were performed on the University of Costa Rica’s Institutional HPC Cluster \footnote{\url{https://hpc.ucr.ac.cr}} using a single Lenovo ThinkSystem SR670 V2 node equipped with 64 Intel Xeon Gold 6338 CPU cores, 1 TB RAM, and one NVIDIA A100 GPU (80 GB). For every reported result, we generated (128{,}000) process simulations and used the corresponding (128{,}000) parameter combinations to train the architectures. Sample generation required approximately 4 hours per network input type \((x_{tj}^{\vee}\) and \(y_{tj}^{\vee})\). Training via Algorithm~\ref{alg:pasquier1} with the BayesFlow library \cite{radev2020bayesflow} took about 2 hours per network. Algorithm~\ref{alg:pasquier2} required roughly 1 hour and 30 minutes to produce (10{,}000) iterations. All computations were performed in Python. For more details, see the repository: \url{https://github.com/luisbarboza27/BayesNetExtremes}.

\section{Application: Analysis of CHIRPS data}\label{sec::aplicacion}

In Costa Rica, most studies of precipitation extremes have used exploratory analyses or classical statistical tools. For example, \cite{GuillenOviedo2020Uniciencia} compares generalized extreme-value (GEV) parameters across 103 meteorological stations in Central America, and regional studies report positive temporal trends in several precipitation- and temperature-based extreme indices \cite{marcela}. More recently, \cite{HugueninSerafinWaylen2023} employed a peaks-over-threshold (POT) framework with a nonstationary point-process model and climate covariates for a subregion smaller than the present study area, using a frequentist fit.

To demonstrate our approach, we analyze daily precipitation intensities at 83 observation sites in the Guanacaste region of Costa Rica using CHIRPS data \cite{funk2015climate,CHIRPS-article}. CHIRPS provides quasi-global daily precipitation (mm) on a 0.05$^\circ$ grid (50$^\circ$S–50$^\circ$N) by merging CHPclim climatologies, satellite retrievals, and in-situ measurements. We restrict attention to September–December (2015–2022) to capture the primary rainy-season peak and limit complex temporal nonstationarities, yielding $n=976$ daily replicates. Pairwise distances among the 83 sites range from 10.9 to 160\,km (mean 57.8\,km), which, under a stationary isotropic exponential correlation, directly governs the decay of spatial dependence. Across sites, zero-precipitation days constitute 51–60\% of records.

\begin{figure}
	\centering
	\begin{tabular}{c}
		\includegraphics[scale=0.5]{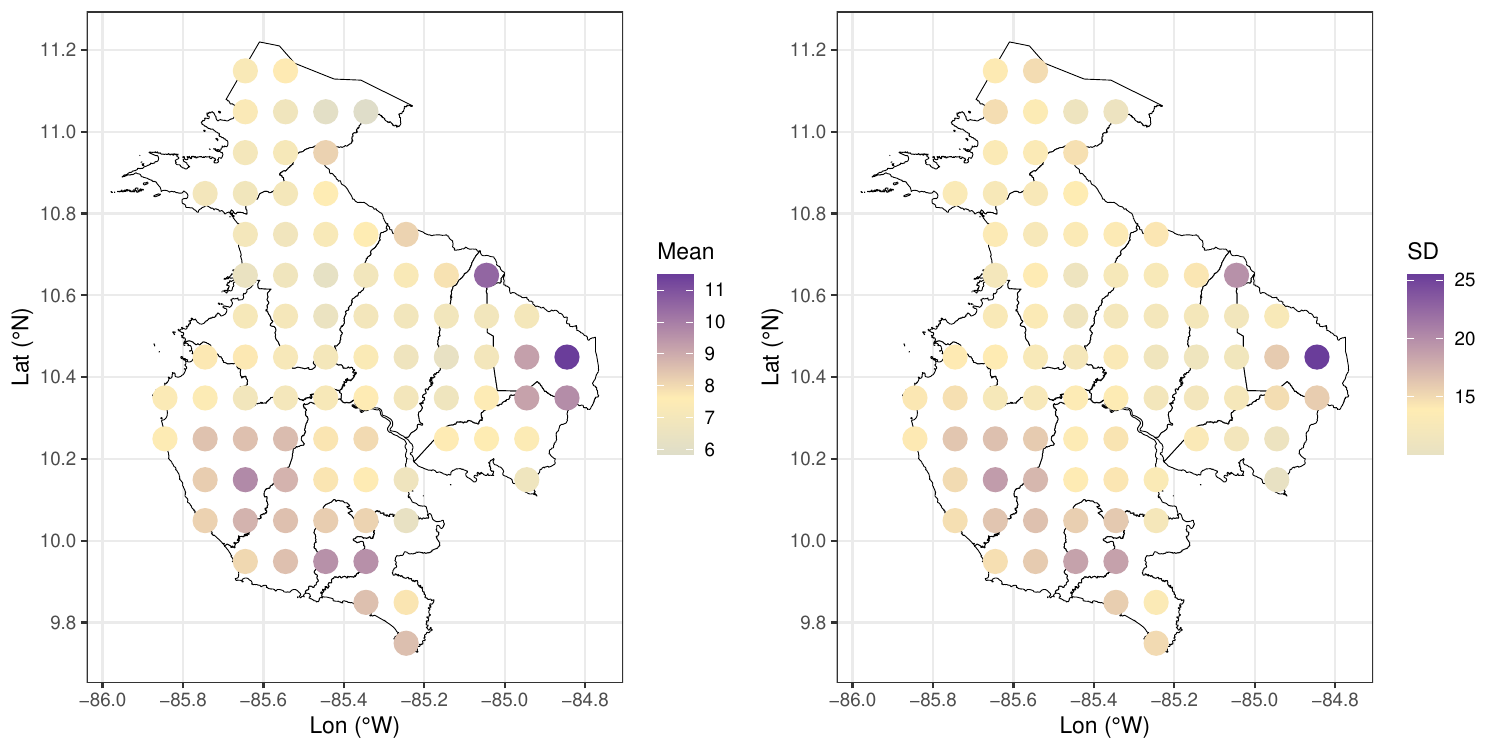}\\
		\includegraphics[scale=0.5]{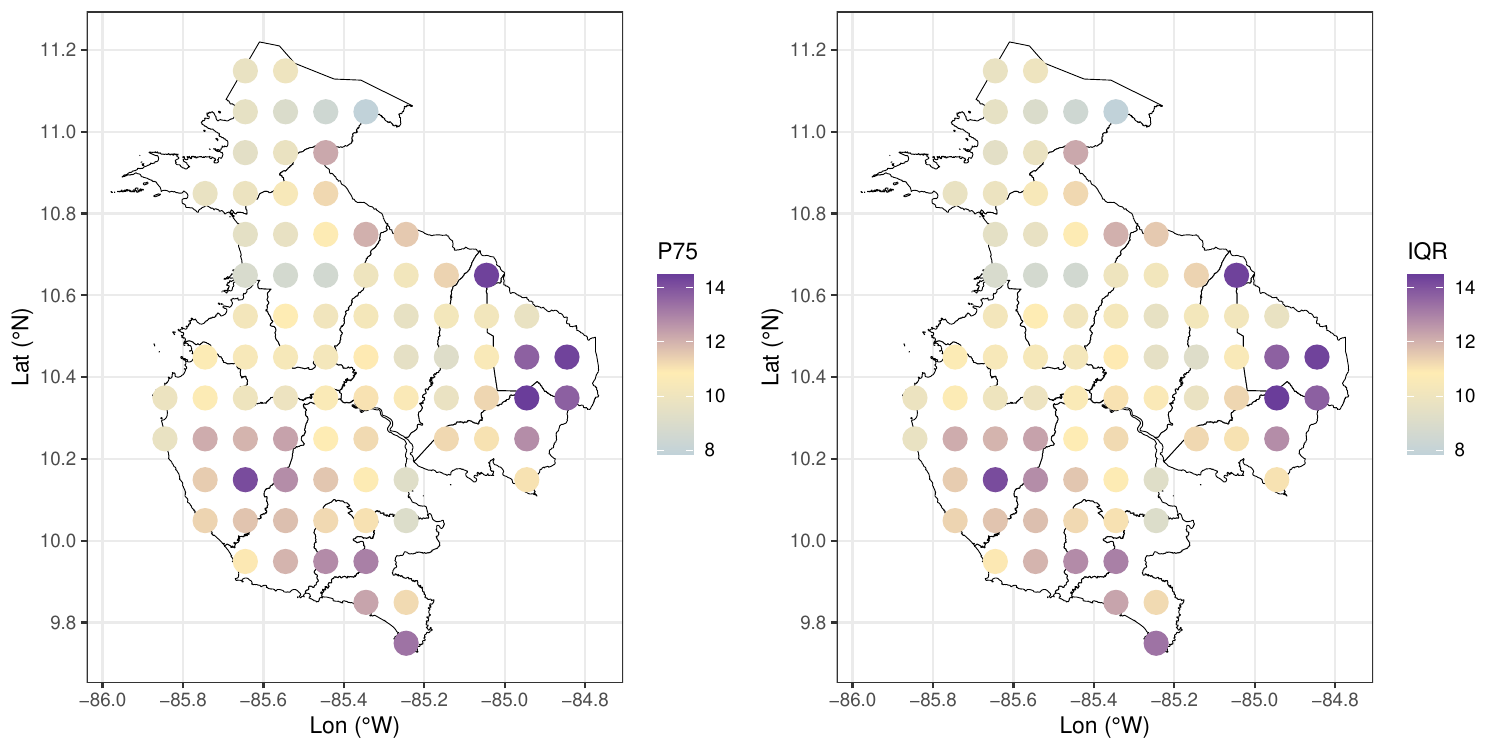}\\
		\includegraphics[scale=0.5]{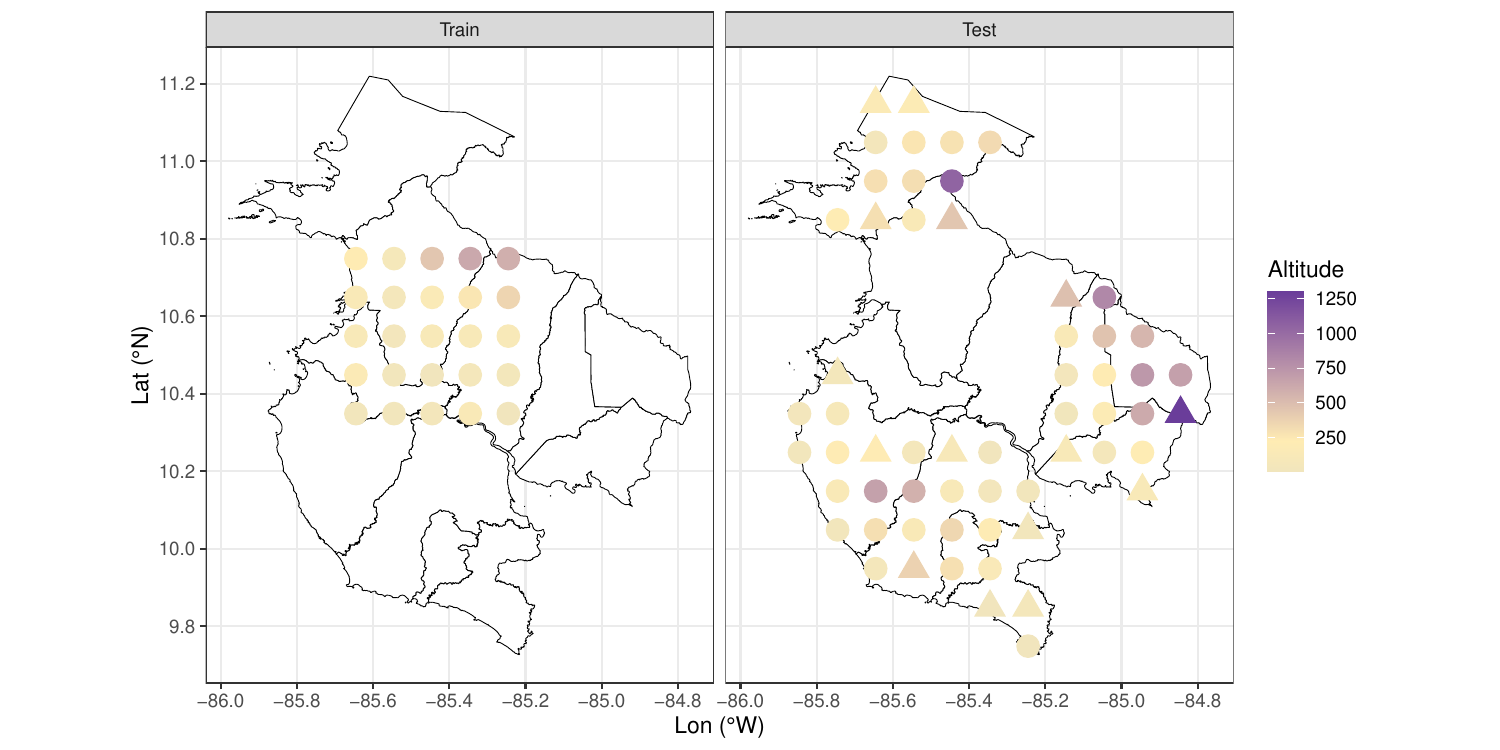}
	\end{tabular}
	\caption{Mean, standard deviation (SD), 75th percentile (P75), and interquartile range (IQR) of precipitation (mm) at each site. Distribution of training and test locations. Sites marked with triangles correspond to those in Figure~\ref{fig:qqplot}.}
	\label{fig:locaciones_media}
\end{figure}

In Figure~\ref{fig:locaciones_media}, we map the mean, standard deviation, 75th percentile, and interquartile range of precipitation at each site, revealing pronounced spatial heterogeneity and systematic variation with latitude and longitude. The figure also delineates the train/test split: we train on $d=25$ sites over 2015–2019 ($n=610$ days) and test on $d=52$ sites over 2020–2022 ($n=366$ days). Because training required a regular subregion, the training domain is relatively small compared with the testing domain.

We apply the spatial-product model in \eqref{eq:modelo_propuesto} to the Guanacaste data and compare eight nested variants (D1–D8) plus the original specification in \cite{yadav2022flexible} (DY). The variants toggle whether the autoregressive factor $X_{2t}^{\mathrm{AR}}$ and the noise factor $X_{1t}$ are spatially constant (superscript “c”) or spatially varying, and whether the spatial-dependence factor $X_{3t}$ is included:
\begin{description}
	\item[D1] $\mathbf{Y}_t=\boldsymbol{\alpha}\,X_{2t}^{\mathrm{AR\text{-}c}}$ (constant AR only).
	\item[D2] $\mathbf{Y}_t=\boldsymbol{\alpha}\,X_{2t}^{\mathrm{AR}}$ (spatially varying AR only).
	\item[D3] $\mathbf{Y}_t=\boldsymbol{\alpha}\,X_{2t}^{\mathrm{AR\text{-}c}}\,X_{3t}$ (constant AR $+$ spatial dependence).
	\item[D4] $\mathbf{Y}_t=\boldsymbol{\alpha}\,X_{2t}^{\mathrm{AR}}\,X_{3t}$ (varying AR $+$ spatial dependence).
	\item[D5] $\mathbf{Y}_t=\boldsymbol{\alpha}\,X_{1t}\,X_{2t}^{\mathrm{AR}}\,X_{3t}$ (varying noise $+$ varying AR $+$ spatial dependence).
	\item[D6] $\mathbf{Y}_t=\boldsymbol{\alpha}\,X_{1t}\,X_{2t}^{\mathrm{AR\text{-}c}}\,X_{3t}$ (varying noise $+$ constant AR $+$ spatial dependence).
	\item[D7] $\mathbf{Y}_t=\boldsymbol{\alpha}\,X_{1t}^{\mathrm{c}}\,X_{2t}^{\mathrm{AR\text{-}c}}\,X_{3t}$ (constant noise $+$ constant AR $+$ spatial dependence).
	\item[D8] $\mathbf{Y}_t=\boldsymbol{\alpha}\,X_{1t}^{\mathrm{c}}\,X_{2t}^{\mathrm{AR}}\,X_{3t}$ (constant noise $+$ varying AR $+$ spatial dependence; main proposal).
	\item[DY] $\mathbf{Y}_t=\boldsymbol{\alpha}\,X_{1t}\,X_{2t}^{\mathrm{c}}\,X_{3t}$ (original formulation in \cite{yadav2022flexible}).
\end{description}

To assess covariate effects on the scale, each D1–D8 variant is embedded in seven log-linear specifications:
\begin{align*}
	\mathbf{M1}:~ & \boldsymbol{\alpha}=\exp(\gamma_0\mathbf{1}_d) \\
	\mathbf{M2}:~ & \boldsymbol{\alpha}=\exp(\gamma_0\mathbf{1}_d+\gamma_{\text{lon}}\mathbf{Z}_1) \\
	\mathbf{M3}:~ & \boldsymbol{\alpha}=\exp(\gamma_0\mathbf{1}_d+\gamma_{\text{lon}}\mathbf{Z}_1+\gamma_{\text{lat}}\mathbf{Z}_2) \\
	\mathbf{M4}:~ & \boldsymbol{\alpha}=\exp(\gamma_0\mathbf{1}_d+\gamma_{\text{lon}}\mathbf{Z}_1+\gamma_{\text{lat}}\mathbf{Z}_2+\gamma_{\text{alt}}\mathbf{Z}_3) \\
	\mathbf{M5}:~ & \boldsymbol{\alpha}=\exp(\gamma_0\mathbf{1}_d+\gamma_{\text{lon}}\mathbf{Z}_1+\gamma_{\text{lat}}\mathbf{Z}_2+\gamma_{\text{alt}}\mathbf{Z}_3+\gamma_{\text{lon}}^2\mathbf{Z}_1^2) \\
	\mathbf{M6}:~ & \boldsymbol{\alpha}=\exp(\gamma_0\mathbf{1}_d+\gamma_{\text{lon}}\mathbf{Z}_1+\gamma_{\text{lat}}\mathbf{Z}_2+\gamma_{\text{alt}}\mathbf{Z}_3+\gamma_{\text{lon}^2}\mathbf{Z}_1^2+\gamma_{\text{lat}^2}\mathbf{Z}_2^2) \\
	\mathbf{M7}:~ & \boldsymbol{\alpha}=\exp(\gamma_0\mathbf{1}_d+\gamma_{\text{lon}}\mathbf{Z}_1+\gamma_{\text{lat}}\mathbf{Z}_2+\gamma_{\text{alt}}\mathbf{Z}_3+\gamma_{\text{lon}^2}\mathbf{Z}_1^2+\gamma_{\text{lat}^2}\mathbf{Z}_2^2+\gamma_{\text{alt}^2}\mathbf{Z}_3^2),
\end{align*}
where $\mathbf{Z}_1,\mathbf{Z}_2,\mathbf{Z}_3$ are standardized longitude, latitude, and altitude. These seven covariate sets (M1–M7) yield $8\times 7=56$ candidate models, plus DY.

For inference, we fix the summary-network architectures at $(n_{\mathrm{LSTM}},n_{\mathrm{Dense}})=(1024,128)$ for D1–D8 and $(1000,2000)$ for DY, guided by our simulation study. We assign weak priors:
$\phi\sim U(-0.85,0.85)$,
$\sigma\sim U(0.05,3)$,
$\beta_3\sim U(2,15)$,
$\rho\sim U(0,2\delta)$ (with $\delta$ the maximum intersite distance),
$\gamma_i\sim N(0,2)$,
and $\beta_1,\beta_2\sim U(0.05,2)$, following \cite{yadav2022flexible}.

\subsection{Results}
Table~\ref{tab:errores} reports MQAE and MQSE (see Section~\ref{sec:metrics}). Model \textbf{D4}--\textbf{M5} attains the lowest MQAE in both training and test sets, followed by \textbf{DY}--\textbf{M5}. For MQSE, \textbf{D4}--\textbf{M5} performs best in training, whereas \textbf{D8}--\textbf{M4} yields the lowest value on the test set, closely followed by \textbf{D4}--\textbf{M5}. Similar magnitudes across training and testing indicate limited overfitting, except for \textbf{M7}, the most complex specification.
\begin{table}
	\centering
	\small
	\begin{tabular}{@{} l c *{7}{c} @{}}
		\toprule
		& & M1 & M2 & M3 & M4 & M5 & M6 & M7 \\
		\midrule
		\multirow{9}{*}{\makecell[l]{\textbf{MQAE}\\\textbf{training}}}
		& D1 &  5.499 &  4.904 &  6.085 &  5.441 &  4.632 &  4.672 &  8.625 \\
		& D2 & 10.640 &  8.228 &  8.766 &  8.369 &  8.296 &  8.682 &  8.511 \\
		& D3 &  7.334 &  5.129 &  4.975 &  5.681 &  5.085 &  5.920 &  4.984 \\
		& D4 &  8.133 &  6.322 &  3.867 &  3.659 & \textcolor{blue}{3.150} &  4.104 &  3.997 \\
		& D5 &  9.687 &  8.262 &  4.130 &  4.121 &  3.563 &  4.113 &  4.137 \\
		& D6 &  8.221 &  6.341 &  4.364 &  4.559 &  4.438 &  3.892 &  4.876 \\
		& D7 &  6.350 &  4.601 &  4.687 &  4.999 &  4.081 &  5.148 &  5.705 \\
		& D8 &  7.716 &  5.990 &  3.756 &  3.451 &  3.206 &  3.804 &  4.040 \\
		& DY &  9.819 &  6.375 &  4.389 &  3.858 &  3.189 &  3.742 &  4.808 \\
		\midrule
		\multirow{9}{*}{\makecell[l]{\textbf{MQAE}\\\textbf{testing}}}
		& D1 &  6.897 &  6.479 &  7.679 &  6.523 &  5.345 &  5.567 &  9.839 \\
		& D2 & 12.649 & 10.150 & 10.587 &  8.991 &  9.048 &  9.287 &  9.870 \\
		& D3 &  9.468 &  7.006 &  6.615 &  7.250 &  6.939 &  8.780 & 75.695 \\
		& D4 & 10.180 &  8.099 &  6.376 &  5.063 & \textcolor{blue}{4.719} &  5.431 & 45.264 \\
		& D5 & 11.725 & 10.220 &  6.252 &  5.535 &  5.338 &  5.873 & 38.634 \\
		& D6 & 10.276 &  8.322 &  6.062 &  5.889 &  5.657 &  5.888 & 91.876 \\
		& D7 &  8.440 &  6.370 &  5.997 &  6.082 &  5.509 &  6.035 & 59.678 \\
		& D8 &  9.749 &  7.863 &  6.016 &  4.823 &  5.128 &  5.264 & 78.180 \\
		& DY & 11.869 &  8.306 &  6.299 &  5.200 &  5.080 &  5.423 & 59.682 \\
		\midrule
		\multirow{9}{*}{\makecell[l]{\textbf{MQSE}\\\textbf{training}}}
		& D1 & 112.557 &  68.384 & 1.06$\times 10^4$ & 336.499 &  44.003 &  44.682 & 414.385 \\
		& D2 & 177.312 & 133.659 &     152.654      & 141.856 & 144.704 & 156.077 & 205.547 \\
		& D3 &  89.494 &  55.599 &      65.627       &  94.351 &  81.976 & 114.847 &  67.256 \\
		& D4 & 101.487 &  63.928 &      33.038       &  31.731 & \textcolor{blue}{28.973} &  36.610 &  40.577 \\
		& D5 & 155.813 & 120.412 &      39.145       &  39.201 &  33.210 &  37.388 &  46.544 \\
		& D6 & 109.488 &  74.898 &      43.442       &  50.467 &  42.769 &  37.642 &  51.600 \\
		& D7 &  64.909 &  44.228 &      63.245       &  55.807 &  39.782 &  51.931 &  58.939 \\
		& D8 &  96.907 &  61.945 &      31.652       &  31.230 &  35.962 &  32.512 &  42.471 \\
		& DY & 162.304 &  81.407 &      45.012       &  40.455 &  30.536 &  33.988 &  53.180 \\
		\midrule
		\multirow{9}{*}{\makecell[l]{\textbf{MQSE}\\\textbf{testing}}}
		& D1 & 162.986 & 121.387 & 1.08$\times 10^4$ & 1256.860 &  78.500 &  79.854 & 292.389 \\
		& D2 & 291.004 & 232.538 &      253.624      &  216.476 & 208.934 & 216.181 & 401.894 \\
		& D3 & 177.721 & 115.473 &      110.437      &  161.368 & 149.288 & 251.766 & 9.80$\times 10^7$ \\
		& D4 & 191.318 & 132.285 &       94.196      &   78.401 & \textcolor{blue}{70.184} &  79.345 & 1.64$\times 10^7$ \\
		& D5 & 264.245 & 215.548 &       98.381      &   83.272 &  77.565 &  88.502 & 9.43$\times 10^6$ \\
		& D6 & 205.347 & 153.852 &       94.735      &   91.270 &  82.238 &  88.254 & 6.19$\times 10^8$ \\
		& D7 & 142.992 &  97.771 &       98.303      &   98.542 &  86.834 &  90.704 & 4.02$\times 10^7$ \\
		& D8 & 185.916 & 132.205 &       88.325      & \textcolor{blue}{69.361} &  89.552 &  73.590 & 2.06$\times 10^8$ \\
		& DY & 269.325 & 157.340 &      102.371      &   76.704 &  72.352 &  77.597 & 5.30$\times 10^8$ \\
		\bottomrule
	\end{tabular}
	\caption{Mean Quantile Absolute Error (MQAE) and Mean Quantile Squared Error (MQSE) for training and test sets. Lower values indicate better performance; the optimal model for each criterion is highlighted in blue.}
	\label{tab:errores}
\end{table}

Some model combinations (e.g., \textbf{D1}–\textbf{M3} and those including $\gamma_{\text{alt}^2}$) exhibit substantially larger errors, likely reflecting greater estimation difficulty due to model complexity or a mismatch between the covariate structure and high-elevation behavior (see Figure~\ref{fig:locaciones_media}).

Figure~\ref{fig:errores} in the appendix presents MQAE by location for training and test sets (test: 2020–2022). Errors are generally uniform except in a small sector of the south and at two eastern sites corresponding to the highest elevations. Although all models include altitude, performance may degrade where few high-elevation stations limit ground-truth constraints for CHIRPS; satellite-dominated estimates carry higher uncertainty that propagates to the model. Expanding training coverage to include more high-elevation sites would likely improve robustness.

\begin{figure}
	\centering
	\includegraphics[scale=0.4]{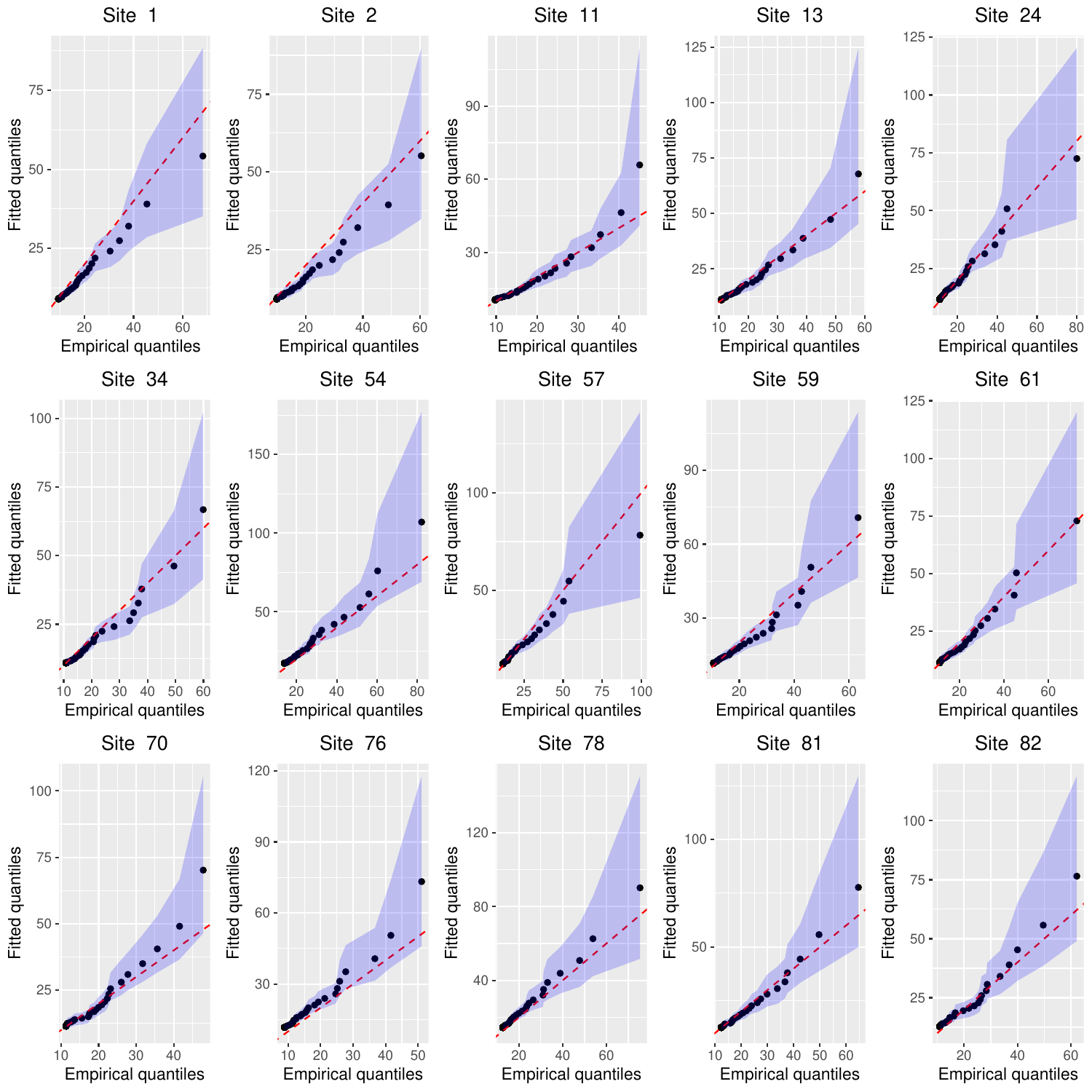}
	\caption{Quantile–quantile plots for the sites indicated by triangles in Figure~\ref{fig:locaciones_media}. Fitted quantiles are simulated from the \textbf{D4}--\textbf{M5} model using posterior mean hyperparameters; shaded bands denote 2.5\% and 97.5\%  quantile simulated uncertainty bands.}
	\label{fig:qqplot}
\end{figure}

In Figure~\ref{fig:qqplot}, quantile–quantile plots for test locations (triangles in Figure~\ref{fig:locaciones_media}) show satisfactory marginal predictive behavior for \textbf{D4} with covariates \textbf{M5}. We therefore select \textbf{D4}--\textbf{M5} as the preferred specification: it delivers the strongest out-of-sample diagnostics, effectively captures spatial heterogeneity in precipitation extremes, and incorporates site-level temporal dependence.

Table~\ref{tab:estadisticos_posterior} summarizes posterior estimates for \textbf{D4}--\textbf{M5}. Only altitude is statistically significant ($\hat{\gamma}_{\mathrm{alt}}=0.25$, 95\% credibility interval $(0.01,0.51)$). We estimate $\phi\approx 0.60$, confirming strong temporal dependence among exceedances above the 75th percentile and highlighting the role of $X_{2t}^{\mathrm{AR}}$. The scale estimate $\sigma\approx 0.13$ indicates non-negligible short-term variability. The tail index $\hat{\xi}=1/2.08\approx 0.48$ suggests heavy tails, consistent with precipitation behavior.
\begin{table}
	\centering
	\begin{tabular}{@{} l *{9}{c} @{}}
		\toprule
		&
		$\gamma_{0}$
		& $\gamma_{\mathrm{lon}}$
		& $\gamma_{\mathrm{lat}}$
		& $\gamma_{\mathrm{alt}}$
		& $\gamma_{\mathrm{lon}^2}$
		& $\phi$
		& $\sigma$
		& $\beta_{3}$
		& $\rho$
		\\
		\midrule
		\makecell[l]{Posterior\\mean}
		&  1.54  &  0.77  & $-$0.14  &  0.25  &  0.82  &  0.60  &  0.13  &  2.08  &  1.67  \\
		\addlinespace
		\makecell[l]{Posterior\\SD}
		&  1.19  &  1.20  &   0.10   &  0.13  &  1.19  &  0.25  &  0.04  &  0.09  &  0.29  \\
		\addlinespace
		\makecell[l]{95\% CI\\lower bound}
		& $-$0.82 & $-$1.47 & $-$0.34 &  0.01  & $-$1.37 & $-$0.09 &  0.04  &  2.00  &  1.17  \\
		\addlinespace
		\makecell[l]{95\% CI\\upper bound}
		&  3.73  &  3.18  &   0.06   &  0.51  &  3.19  &  0.88  &  0.20  &  2.32  &  2.32  \\
		\addlinespace
		ESS/min
		& 107 &  98  &  99   & 105 & 108 & 101 & 109 & 116 & 114 \\
		\bottomrule
	\end{tabular}
	\caption{Posterior summary statistics for the \textbf{D4}--\textbf{M5} model.}
	\label{tab:estadisticos_posterior}
\end{table}

Finally, Figure~\ref{fig:retornos} (appendix) presents return levels across multiple periods under the preferred model. The largest return levels occur over mountainous areas of Guanacaste and the southern Nicoya Peninsula, consistent with prior flood-risk evidence in Costa Rica \cite{QuesadaRoman2022EnvSciPolicy,HugueninSerafinWaylen2023}.

\section{Conclusions}\label{sec::conclusiones}

We extend the Bayesian factor model of \cite{yadav2022flexible} by adding a temporal autoregressive component that captures residual dependence in threshold exceedances. This additional factor can be specified as spatially varying or spatially constant, thereby increasing flexibility for modeling extremes across heterogeneous settings. The framework also accommodates relevant covariates—such as latitude, longitude, altitude, and other domain-specific variables—and explicitly quantifies predictive uncertainty for rare, high-impact events.

We develop a Gibbs sampler that leverages Bayesian neural network architectures to avoid the computational burden of the full censored likelihood. This strategy accelerates inference in high-dimensional parameter spaces, preserves heavy-tail behavior, and captures temporal dependence without sacrificing accuracy. Although training the networks demands substantial computational resources, once trained, the estimator produces posterior inferences rapidly for similar test datasets. The estimation methodology relies entirely on simulation from the hierarchical model components, aligning well with the structure proposed by \cite{yadav2022flexible} and with the extensions introduced here. In complementary simulation studies, despite known challenges in estimating the copula parameters for the $X_3$ component, we observed improved MCMC mixing and estimation accuracy when using moderately more complex architectures; \cite{yadav2022flexible} reports similar difficulties.

The proposed Gibbs scheme generalizes beyond the present extreme-value application. Whenever the parameter vector can be partitioned into blocks with similar inferential characteristics, one can obtain approximate conditional posteriors for each block via amortized methods such as BayesFlow or, alternatively, approximate Bayesian computation, and then interleave these conditionals within a Gibbs routine to produce joint posterior samples.

To illustrate the methodology, we evaluated a suite of nested models for precipitation extremes in Guanacaste, Costa Rica, and selected the factor combination that performed best on out-of-sample diagnostics. The chosen specification elucidates spatial patterns of return periods at multiple time horizons and provides a decision-support tool for infrastructure planning and climate risk management. This contribution is particularly salient in Guanacaste, one of Costa Rica’s most environmentally sensitive regions: its tropical dry climate exhibits pronounced oscillations between drought and intense precipitation, it is directly influenced by ENSO from the Pacific and indirectly by Caribbean wave and tropical-cyclone activity, and it features vulnerable infrastructure and distinctive geological conditions. Natural disasters in the region trigger direct impacts—flooding, crop failures, population displacement—and indirect consequences, including escalating infrastructure and insurance costs \cite{costa_rica_impacto_2019}. By improving the characterization of extremes, our study delivers actionable information to support local planning and climate-adaptation strategies.

\bibliographystyle{unsrt}  
\bibliography{references}  

\appendix

\section{Supplementary Information}

\begin{figure}[h!]
	\centering
	\includegraphics[scale=0.63]{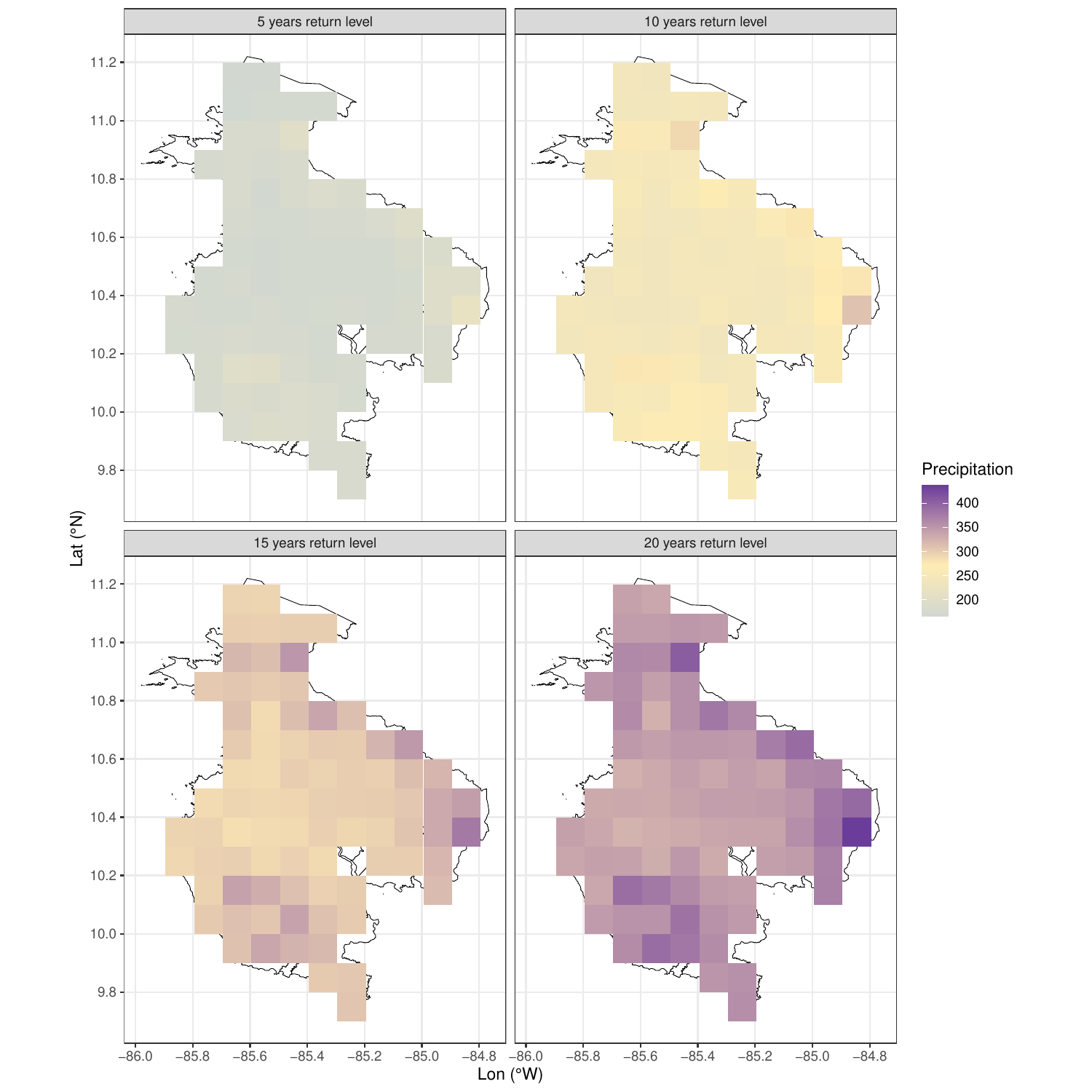}
	\caption{Precipitation return levels (mm) for multiple return periods, estimated using the \textbf{D4}–\textbf{M5} model.}
	\label{fig:retornos}
\end{figure}

\begin{algorithm}
	\caption{Amortized Bayesian Inference via the BayesFlow Method \cite{radev2023jana}}\label{alg:bayesflow}
	\begin{algorithmic}[1]
		\State \textbf{Training Phase} (online learning with batch size $M$):
		\Repeat
		\For{$m = 1,\dots,M$}
		\State Sample model parameters from the prior: $\boldsymbol{\Theta}^{(m)}\sim p(\boldsymbol{\Theta})$
		\For{$i = 1,\dots,n$}
		\State Sample noise instance: $\xi^i\sim p(\xi)$
		\State Simulate synthetic observation: $\boldsymbol{y}_i^{(m)} = g(\boldsymbol{\Theta}^{(m)},\xi^i)$
		\EndFor
		\State Compute summary statistics: $\tilde{\boldsymbol{y}}^{(m)} = h_{\psi}(\boldsymbol{y}_{1:n}^{(m)})$
		\State Forward‐pass through inference network: $\boldsymbol{w}^{(m)} = f_{\phi}(\boldsymbol{\Theta}^{(m)};\,\tilde{\boldsymbol{y}}^{(m)})$
		\EndFor
		\State Compute the loss according to \eqref{eq:montecarlo_completo} using the batch $\{(\boldsymbol{\Theta}^{(m)},\tilde{\boldsymbol{y}}^{(m)},\boldsymbol{w}^{(m)})\}_{m=1}^M$
		\State Update network parameters $\phi,\psi$ via backpropagation
		\Until{convergence to \(\hat\phi,\hat\psi\)}
		\vspace{0.5em}
		\State \textbf{Inference Phase} (given observed or test data \(\boldsymbol{y}^o_{1:n}\)):
		\State Compute summary statistics: \(\tilde{\boldsymbol{y}}^o = h_{\hat\psi}(\boldsymbol{y}^o_{1:n})\)
		\For{$l = 1,\dots,L$}
		\State Sample latent code: \(\boldsymbol{w}^{(l)}\sim\mathcal{N}_D(0,I)\)
		\State Invert through inference network: \(\boldsymbol{\Theta}^{(l)} = f^{-1}_{\hat\phi}(\boldsymbol{w}^{(l)};\,\tilde{\boldsymbol{y}}^o)\)
		\EndFor
		\State \Return \(\{\boldsymbol{\Theta}^{(l)}\}_{l=1}^L\) as samples from \(p(\boldsymbol{\Theta}\mid\boldsymbol{y}^o_{1:n})\)
	\end{algorithmic}
\end{algorithm}

\begin{table}
	\centering
\begin{tabular}{@{}lcccc@{}}
	\toprule
	Layer & Input size & Output size & Filter size & Number of filters \\
	\midrule
	Input & $[n, d{+}4]$ & — & — & — \\
	LSTM & $[n, d{+}4]$ & $[n, n_{\text{LSTM}}]$ & — & — \\
	LSTM & $[n, n_{\text{LSTM}}]$ & $[n_{\text{LSTM}}]$ & — & — \\
	Dense (ReLU) & $[n_{\text{LSTM}}]$ & $[n_{\text{Dense}}]$ & — & — \\
	Dense (ELU) & $[n_{\text{Dense}}]$ & $[n_{\text{Dense}}]$ & — & — \\
	\midrule
	Input & $[n, d_1, d_2, 1]$ & — & — & — \\
	TimeDistributed Conv2D & $[n, d_1, d_2, 1]$ & $[n, d_1, d_2, 32]$ & $3 \times 3$ & $32$ \\
	TimeDistributed Conv2D & $[n, d_1, d_2, 32]$ & $[n, d_1, d_2, 64]$ & $3 \times 3$ & $64$ \\
	TimeDistributed Flatten & $[n, d_1, d_2, 64]$ & $[n, d_1 \cdot d_2 \cdot 64]$ & — & — \\
	LSTM & $[n, d_1 \cdot d_2 \cdot 64]$ & $[n_{\text{LSTM}}]$ & — & — \\
	LSTM & $[n, n_{\text{LSTM}}]$ & $[n_{\text{LSTM}}]$ & — & — \\
	Dense (ReLU) & $[n_{\text{Dense}}]$ & $[n_{\text{Dense}}]$ & — & — \\
	Dense (ELU) & $[n_{\text{Dense}}]$ & $[n_{\text{Dense}}]$ & — & — \\
	\bottomrule
\end{tabular}
\caption{Architectural Specification of $R_{\alpha}$ and $R_{\mathbf{X}}$ Summary Networks.}
\label{tab::archiqs}
\end{table}

\begin{figure}[h!]
	\centering
	\includegraphics[width=\textwidth]{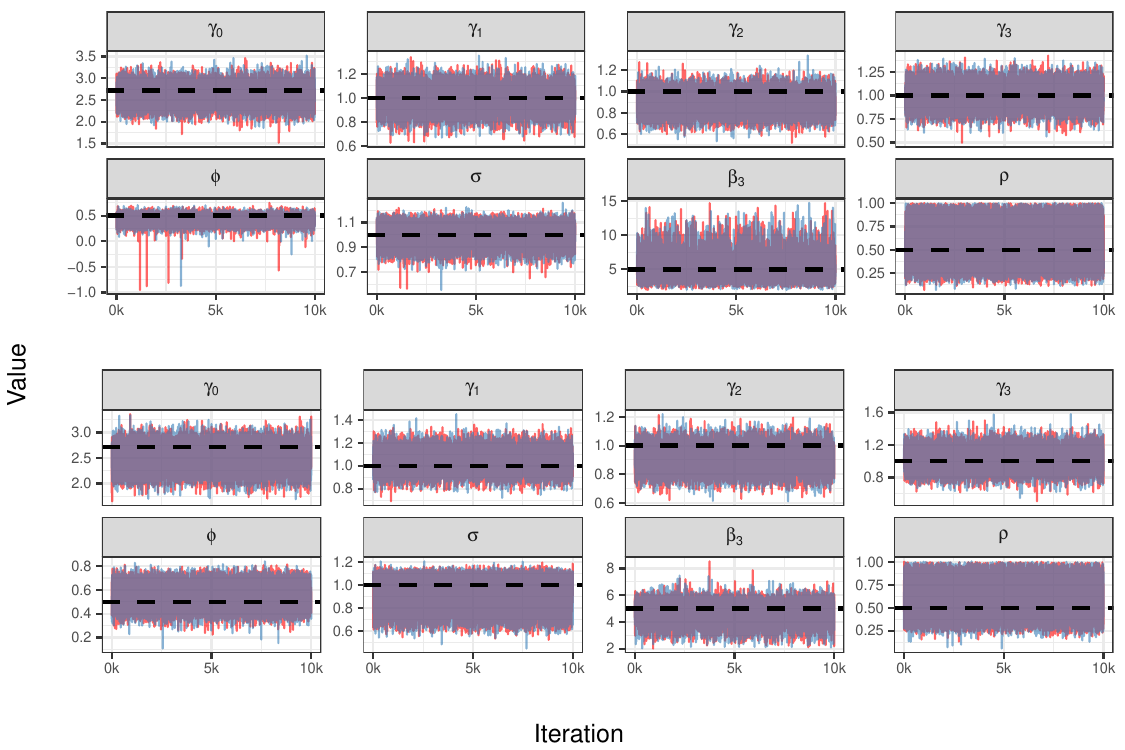}
	\caption{
		Posterior trace plots for the simulation‐study hyperparameters. The first two rows correspond to Scenario 1 and the last two rows to Scenario 5. Each panel shows two chains (red and blue) initialized at different starting values. We ran 10000 MCMC iterations, and the vertical black lines mark the true parameter values.
	}
	\label{fig:traceplot_simulacion}
\end{figure}

\begin{figure}[h!]
	\centering
	\includegraphics[width=\textwidth]{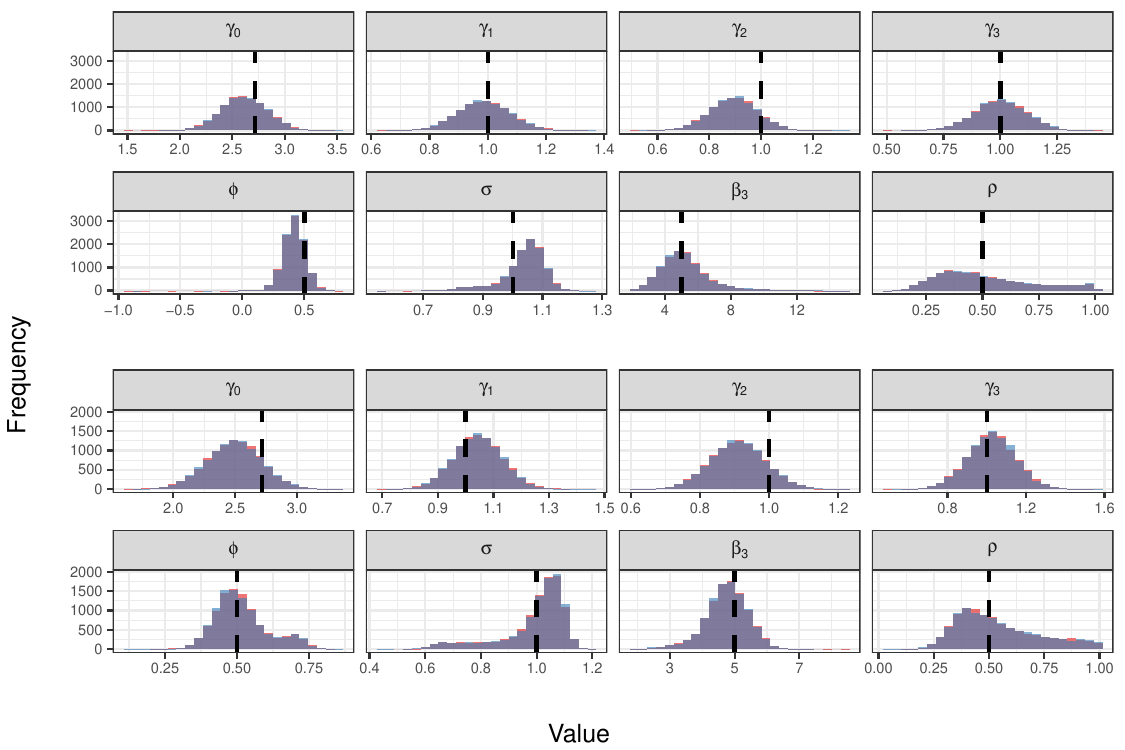}
	\caption{
		Histogram of posterior samples for the simulation‐study hyperparameters. The top two rows correspond to Scenario 1 and the bottom two to Scenario 5. Each histogram overlays two chains (red and blue) initialized with different starting values. We ran 10000 MCMC iterations, and the horizontal black lines mark the true parameter values.
	}
	\label{fig:histograma_simulacion}
\end{figure}

\begin{figure}[h!]
	\centering
	\includegraphics[scale=0.65]{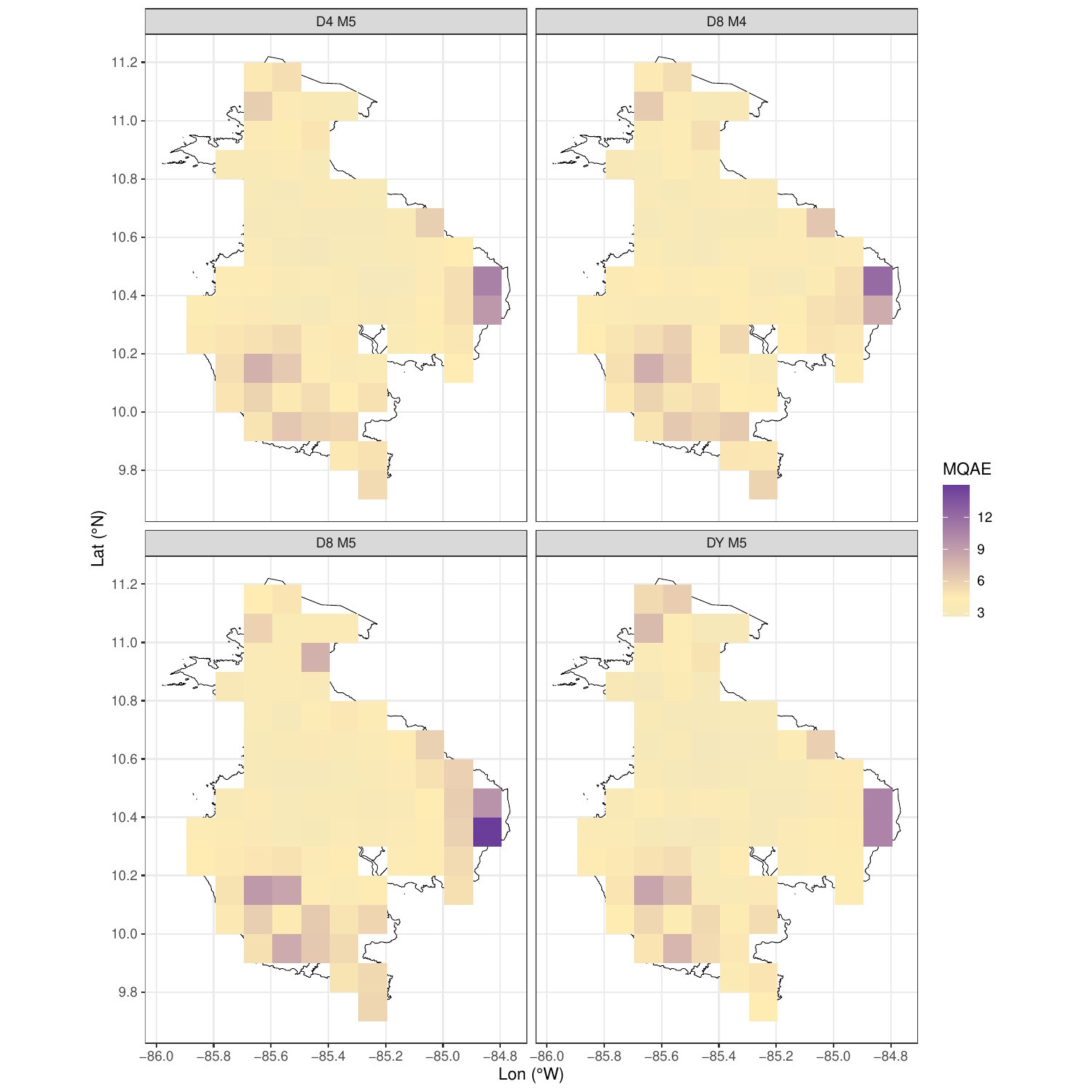}
	\caption{Mean Quantile Absolute Error (MQAE) of precipitation (mm) by location using the years $2020$, $2021$, and $2022$ for various model combinations.}
	\label{fig:errores}
\end{figure}


\end{document}